\documentclass[11pt]{amsart}
\usepackage[a4paper,top=80pt,lmargin=80pt,rmargin=80pt,bottom=80pt]{geometry}

\usepackage{graphicx}

\usepackage[]{cite}

\usepackage{amsmath}
\usepackage[]{amssymb}
\usepackage[]{mathrsfs}
\usepackage[]{eucal}
\usepackage[]{tensor}
\usepackage[]{amsthm}
\usepackage[]{bm}

\DeclareMathAlphabet{\mathcalO}{OMS}{cmsy}{m}{n} 

\usepackage{booktabs}
\usepackage{multirow}

\usepackage{tikz}
\usepackage{tikz-cd}
\newcommand{\btkz}{\begin{tikzpicture}}
\newcommand{\etkz}{\end{tikzpicture}}

\newcommand{\df}{\mathrm{d}}
\newcommand{\cl}[1]{\overline{#1}}

\newcommand{\qqd}{\, , \quad}
\newcommand{\bc}{\begin{center}}
\newcommand{\ec}{\end{center}}
\newcommand{\be}{\begin{equation}}
\newcommand{\ee}{\end{equation}}
\newcommand{\benn}{\begin{equation*}}
\newcommand{\eenn}{\end{equation*}}

\DeclareMathOperator\supp{supp}

\newcommand{\veps}{\varepsilon}

\newcommand{\lb}{\left<}
\newcommand{\rb}{\right>}
\newcommand{\norm}[1]{\left\lVert #1 \right\rVert}

\newcommand{\defeq}{\mathrel{\mathop:}=}

\newcommand{\Ran}{\mathrm{Ran}}
\newcommand{\Ker}{\mathrm{Ker}}

\newcommand{\Ree}{\mathrm{Re}}
\newcommand{\Imm}{\mathrm{Im}}

\newcommand{\nn}{\mathbb{N}}
\newcommand{\zz}{\mathbb{Z}}
\newcommand{\qq}{\mathbb{Q}}
\newcommand{\rr}{\mathbb{R}}
\newcommand{\cc}{\mathbb{C}}
\newcommand{\0}{\emptyset}

\newcommand{\hilb}[1]{\mathcalO{#1}} 
\newcommand{\HH}{\hilb{H}}

\newcommand{\Can}{\mathcalO{C}}

\usepackage[unicode]{hyperref}
\usepackage{xcolor}
\definecolor{pastgreen}{HTML}{669900}
\definecolor{pastblue}{HTML}{336699}
\definecolor{pastred}{HTML}{990000}
\definecolor{linkcol}{HTML}{663333}
\definecolor{graya}{rgb}{0.95,0.95,0.95}
\hypersetup{colorlinks,linkcolor={pastblue},citecolor={pastblue},urlcolor={pastblue}}

\theoremstyle{plain} \newtheorem{tm}{Theorem}[section]
\theoremstyle{plain} \newtheorem{lm}[tm]{Lemma}
\theoremstyle{plain} \newtheorem{defn}[tm]{Definition}
\newcommand{\btm}{\begin{tm}}
\newcommand{\etm}{\end{tm}}
\newcommand{\blm}{\begin{lm}}
\newcommand{\elm}{\end{lm}}
\newcommand{\bdefn}{\begin{defn}}
\newcommand{\edefn}{\end{defn}}

\begin{document}

\begin{flushright}
\texttt{ZTF-EP-26-14}
\end{flushright}

\vspace{30pt}

\title{Spectral measures and sharpness of quantum speed limits}

\author[I. Smoli\'c]{Ivica Smoli\'c}
\address[]{University of Zagreb Faculty of Science, Department of Physics, Bijeni\v cka cesta 32, 10000 Zagreb, Croatia}
\email{ismolic@phy.hr}


\begin{abstract}
Quantum speed limits bound the time required for a given state to evolve into a given subspace of the Hilbert space under specific energy constraints. We prove that for several standard orthogonalization time bounds, the unique spectral measure associated with the initial vector that saturates the bound is a sum of two Dirac measures. Furthermore, we constrain the families of spectral measures that come arbitrarily close to these bounds and construct examples with measures which are purely absolutely continuous and purely singular continuous with respect to the Lebesgue measure. As a complementary perspective, we discuss the geometric intuition behind quantum speed limits provided by the numerical range of the evolution operator.
\end{abstract}

\maketitle

\section{Introduction} 

Quantum physics is shaped by a celebrated class of results, including the uncertainty principle, spin-statistics theorem, and no-cloning theorems, which delineate the states and processes that are physically possible. Among these, quantum speed limits (QSL) establish fundamental bounds on the rate at which quantum states, either pure or mixed, can evolve to some final state, given energy constraints. In particular, so-called \emph{orthogonalization time} is a minimal time necessary for a pure state to evolve into an orthogonal state.

\smallskip

An early example of a QSL was given by Mandelstam and Tamm \cite{MT45} in 1945, in attempt to gain conceptual understanding of the elusive time-energy uncertainty relation\footnote{For further discussion about common misinterpretations and the physical meaning of the time-energy uncertainty relation, see \cite{AB61,Fock62,AB64,UH85,GV05}.}. Mandelstam--Tamm (TM) bound establishes a lower bound on the orthogonalization time in terms of the reciprocal energy uncertainty (standard deviation of the Hamiltonian) of the initial state. This result was later generalized to time-dependent Hamiltonians \cite{AA90,Pfeifer93}, mixed states \cite{PF95,IL21,HAS22} and evolution of a whole subspace \cite{AM22}. Another well-known QSL is the Margolus--Levitin (ML) bound \cite{ML98}, which establishes a lower bound on the orthogonalization time in terms of the average energy of the initial state. The ML bound was generalized by Luo and Zhang \cite{LZ05} to a family of inequalities for each energy moment. While some of the aforementioned references also discuss extensions of MT and ML bounds under the relaxed condition in which the fidelity between the initial and final states is nonzero, a quest for sharp bounds of this type were first initiated by the numerical work in \cite{GLM03} and later analytically treated in \cite{HS23ml} (for additional clarifications and refinements see \cite{NAS22,ABLK23,Chau23,CZ24}). A detailed overview of the QSL results can be found in \cite{Frey16,DC17}, while \cite{QU26} offers a complementary survey of uncertainty principles in quantum physics.

\smallskip

Implications of the QSLs were discussed in context of the ultimate computation limits \cite{ML98,Lloyd00} (suggesting an estimate of maximal number of logical operations per second at the order of $\sim\!\!E/\hbar$ for the given energy resources $E$), neutrino oscillations \cite{BJS19,BJB25} and analogous speed limits in classical systems \cite{OO18,SCMdC18,PCD21,Hamazaki22,GNGdCG22}. QSL bounds have been experimentally tested using quantum photonic states in superconducting resonators \cite{Wu24}, few- and many-body states in a programmable superconducting quantum processor \cite{Zhu25}, and nuclear spins in chloroform molecules \cite{Pires24}.

\smallskip

The present work focuses on the saturation of QSLs, uniqueness of the saturating states, and the structure of near-saturating states. While some of the early works (e.g.~\cite{Pfeifer93,PF95,ML98}) note the MT and ML bound can be saturated in a simple two-level quantum system, first proofs of uniqueness of the saturating state have appeared in 1998, for the MT bound \cite{HM98} (using combination of geometric and algebraic tools) and ML bound \cite{SBTT98} (using Lagrange multipliers). A more systematic approach to QSLs with auxiliary nonnegative functions, introduced in \cite{ML98} and generalized in \cite{LZ05}, was efficiently used in analysis of uniqueness of saturating states for ML bound \cite{KZ06}, LZ bound \cite{ZZ06} and MT bound \cite{LT09} (a more recent uniqueness result \cite{Sonn26} is focused on finite-dimensional spaces only). However, by tacitly assuming a purely point spectrum for the Hamiltonian, some of these papers \cite{HM98,SBTT98,LT09} exclude even elementary systems such as the hydrogen atom, thereby restricting the scope of their conclusions. Other works \cite{LZ05,KZ06,ZZ06} invoke the spectral theorem to accommodate more general spectral types, but the corresponding arguments leave many technical points implicit. Moreover, to the best of our knowledge, none of these papers addresses how closely the QSL bounds can be approached using states belonging to absolutely continuous or singular continuous spectral subspaces. Our aim is to address these questions at the level of functional-analytic rigor and to clarify the numerous subtleties involved.

\medskip

\textbf{Outline}. The paper is organized as follows. In Section 2 we define the central object of our study, orthogonalization time for evolving quantum states, introduce the key elements of the spectral theory and discuss behaviour of handful of elementary, but important examples. In Section 3 we revisit several known theorems, providing rigorous and general formulations that make explicit the underlying assumptions and extend beyond the finite-dimensional and pure-point-spectrum settings commonly considered in the literature. In Section 4 we prove that the state which saturates bounds proven in the Section 3 exist and has unique associated spectral measure (while uniqueness of the state vector depends on the dimension of the eigenspaces). In Section 5 we constraint form of the ``near-saturating'' states and construct examples with absolutely continuous and singular continuous spectral measures. In Section 6 we develop an approach to QSLs from the perspective of numerical range of operators, while in Section 7 we briefly discuss remaining open questions. At the end of the paper we present three appendices: Appendix A with proofs of several auxiliary inequalities, used in the main text, Appendix B in which we reconstruct form of spectral measures from a given integral constraint, and Appendix C with the construction and analysis of several basic properties of the Cantor measure.

\smallskip

\textbf{Notation and conventions}. A set difference is denoted by the minus sign, $A - B = \{x \in A : x \notin B \}$. Characteristic function for a set $S$ is denoted by $1_S$. Dirac measure on $\rr$, supported at point $a \in \rr$, is defined as
\be
\delta_a(S) = \left\{ \begin{array}{rl} 1 \, , & a \in S \\ 0 \, , & a \notin S \end{array} \right. 
\ee
for any subset $S \subseteq \rr$. The inner product $\lb \,.\,,\,.\, \rb$ in the Hilbert space is assumed to be conjugate linear in the first variable and linear in the second (``physicists' convention''). Domain of an operator $A$ is denoted by $D(A)$. Identity operator is denoted by $I$. For simplicity, we choose physical units in which $\hbar = 1$.

\bigskip

\section{Orthogonalization time} 

We shall study time evolution of a pure quantum state in a closed system. Dynamics on a Hilbert space $\HH$ is generated by a time-independent Hamiltonian $H$, which is not necessarily bounded operator. Our goal is to determine the bounds on earliest time at which the evolved state $\exp(-iHt) u$ becomes orthogonal to its initial state $u$, classify states which saturate these bounds and find how closely may non-saturating states approach these bounds.

\bdefn
Let $\HH$ be a complex separable Hilbert space, with $\dim\HH \ge 2$, and let $H : D(H) \to \HH$ be a densely defined, self-adjoint Hamiltonian. Then for each unit $u \in \HH$ we define the orthogonalization time
\be
t_\perp(u) \defeq \inf \left\{ t > 0 : \, \big< u, e^{-iHt} u \big> = 0 \right\} ,
\ee
with the convention that $\inf\0 = +\infty$.
\edefn

\noindent
As the Hamiltonian $H$ is a self-adjoint operator, its spectrum is a closed subset of the real numbers, $\sigma(H) \subseteq \rr$, and by the spectral theorem \cite{RS1,HallQ,Teschl} there is a unique projection-valued spectral measure $P_H$, a map from the Borel $\sigma$-algebra on $\rr$ to the set of orthogonal projections on $\HH$, such that formally
\be
H = \int_{\sigma(H)} \omega \, \df P_H(\omega) .
\ee
Moreover, for each vector $u \in \HH$ there is the associated (scalar) spectral measure\footnote{For simplicity, we suppress the additional ``$H$'' index on measure $\mu_u$.} $\mu_u$, a Borel measure on $\rr$, such that $\mu_u(B) = \lb u, P_H(B)u \rb$ for each Borel set $B \subseteq \rr$. Property $P_H(\rr) = I$ implies that $\mu_u$ is a finite measure, $\mu_u(\rr) = \norm{u}^2 < \infty$. In particular, for a unit vector, $\norm{u} = 1$, it is a probability measure, $\mu_u(\rr) = 1$. 

\smallskip

More generally, if $g : \rr \to \cc$ is a Borel function, then the functional calculus allows us (see e.g.~Section 3.1 in \cite{Teschl} and Section 5.3 in \cite{Schm}) to define self-adjoint operator
\be\label{eq:funcalc}
g(H) = \int_{\sigma(H)} g(\omega) \, \df P_H(\omega) ,
\ee
on a dense domain
\be
D(g(H)) = \left\{ u \in \HH : \int_{\sigma(H)} |g(\omega)|^2 \, \df\mu_u(\omega) < \infty \right\} .
\ee
Operator (\ref{eq:funcalc}) satisfies
\be\label{eq:g(H)}
\lb v, g(H) v \rb = \int_{\sigma(H)} g(\omega) \, \df\mu_v(\omega)
\ee
for any $v \in D(g(H))$. Thus, the time evolution operator $\exp(-iHt) : \HH \to \HH$ is a well-defined (unitary) operator for each time $t \in \rr$.

\smallskip

The analysis of the orthogonalization time is focused on the function $f_u : \rr \to \rr$,
\be
f_u(t) \defeq \big< u, e^{-iHt} u \big> .
\ee
Again, using the spectral theorem and functional calculus, we may write it as a convenient integral
\be\label{eq:fu}
f_u(t) = \int_{\sigma(H)} e^{-i\omega t} \, \df\mu_u(\omega) .
\ee
We immediately see\footnote{Note that $\supp \mu_u \subseteq \sigma(H)$, so that $\mu_u(\rr - \sigma(H)) = 0$.} that the function $f_u$ is, up to rescaling and possible normalization choice, equal to the Fourier transform $\hat{\mu}_u$ of the spectral measure. There is a substantial literature devoted to study of the decay properties of the function $f_u(t)$, starting from classical Riemann--Lebesgue lemma, over Wiener's theorem and RAGE theorem \cite{RS3,Teschl}, to later refinements \cite{AS81,KPG92,Last96} (for a comprehensive review, see \cite{MW13}). These results, however, do not inform us about the central problem of this work, infimum of the set of zeros of $f_u(t)$.

\smallskip

Integral in (\ref{eq:fu}) has three prominent types of contributions. If $\lambda$ is the Lebesgue measure on $\rr$, by the standard decomposition results in measure theory \cite{RS1,Borthwick} any finite Borel measure $\mu$ on $\rr$ may be uniquely written as a sum of absolutely continuous, singular continuous, and pure point parts,
\be
\mu = \mu_\mathrm{ac} + \mu_\mathrm{sc} + \mu_\mathrm{pp} ,
\ee
with respect to the measure $\lambda$. The absolutely continuous part has a property that $\mu_\mathrm{ac}(B) = 0$ for all Borel sets $B \subseteq \rr$ which have zero Lebesgue measure, $\lambda(B) = 0$. The singular continuous part $\mu_\mathrm{sc}$ is supported on some zero Lebesgue measure set, but $\mu_\mathrm{sc}(\{x\}) = 0$ for all $x \in \rr$. The pure point is a sum of Dirac measures, $\mu_\mathrm{pp} = \sum_{x \in S} \delta_x$ for some countable set $S \subseteq \rr$. We may apply this decomposition to the spectral measure $\mu_u$ and, analogously, the whole Hilbert space splits as $\HH = \HH_\mathrm{ac} \oplus \HH_\mathrm{sc} \oplus \HH_\mathrm{pp}$, where $\HH_{xx}$ for $xx \in \{\mathrm{ac},\mathrm{sc},\mathrm{pp}\}$ consists of vectors $u$ such that $\mu_u$ has only $xx$ component (see e.g.~Lemma 3.18 in \cite{Teschl}). Similarly, spectrum $\sigma(H)$ is a union of disjoint subsets $\sigma_{xx}(H) = \sigma(H|_{\HH_{xx}})$.

\smallskip

Examples involving any of these components of the spectral measure abound in physics \cite{RS1,Teschl,Borthwick}. For a free particle with Hamiltonian $H = -\Delta$, the spectral measure $\mu_u$ associated with any state (wave package) $u \in L^2(\rr^3)$ is absolutely continuous with respect to Lebesgue measure. In contrast, for the harmonic oscillator $H = -\Delta + |x|^2$, the spectral measure $\mu_v$ associated with any state $v \in L^2(\rr^3)$ is pure point. For the hydrogen atom, with Hamiltonian $H = -\Delta - |x|^{-1}$, the spectral measure of a state may contain both an absolutely continuous component, corresponding to unbound states, and a pure point component, corresponding to bound states. Finally, the almost Mathieu operator $H_{\lambda,\alpha,\theta}$ on $\ell^2(\zz)$, which models an electron on a lattice immersed in a magnetic field, has purely singular continuous spectrum\footnote{The fact that its spectrum is a Cantor set for all irrational $\alpha$ and for all $\lambda \ne 0$ is the famous ``Ten Martini Problem'' \cite{Simon82}, solved by Avila and Jitomirskaya in 2009 \cite{AJ09}.} at the critical coupling $|\lambda| = 1$, for irrational $\alpha$ (magnetic flux per a unit cell, measured in units of the flux quantum) and Lebesgue a.e.~phase $\theta$ \cite{LastSpectral,LastExotic}.

\smallskip

Orthogonalization time $t_\perp(u)$ may be infinite if, for example, $u$ is an eigenvector of the Hamiltonian, $H u = E u$ for some $E \in \rr$. In this case the time evolution is merely a phase change of the initial vector, $t \mapsto e^{-iEt} u$, so that $f_u(t) = e^{-iEt}$ has no zeros and $t_\perp(u) = \infty$. A bit less trivial example can be constructed if the Hamiltonian has at least two orthonormal eigenvectors, $e_1,e_2 \in D(H)$, such that $H e_i = E_i e_i$ for $i \in \{1,2\}$ and $E_1 \le E_2$. Then we introduce the initial vector
\be
u = c_1 e_1 + c_2 e_2
\ee
with $c_1,c_2 \in \cc^\times$, such that $|c_1|^2 + |c_2|^2 = 1$. Here we have
\be
f_u(t) = |c_1|^2 e^{-iE_1 t} + |c_2|^2 e^{-iE_2 t} ,
\ee
and, by the triangle inequality, $|f_u(t)|^2 \ge \big| |c_1|^2 - |c_2|^2 \big|$. If $|c_1| \ne |c_2|$, the function $f_u$ has no zeros and $t_\perp(u) = \infty$. On the other hand, if $|c_1| = |c_2|$, normalization fixes $|c_1| = |c_2| = 1/\sqrt{2}$, from where we get
\be
f_u(t) = \frac{1}{2} \, \big( 1 + e^{-i(E_2 - E_1)t} \big) e^{-iE_1 t} .
\ee
The degenerate case, in which $E_1 = E_2$, again reduces to $f_u(t) = e^{-iE_1 t}$ with $t_\perp(u) = \infty$. On the other hand, if the eigenvectors are nondegenerate, $E_1 \ne E_2$, orthogonalization time is finite, $t_\perp(u) = \pi/(E_2 - E_1)$. We stress that $t_\perp(u)$ does not depend on the phases of coefficients $c_1$ and $c_2$, that is on the relative phase of eigenvectors $e_1$ and $e_2$. This simple analysis holds an important lesson: Orthogonalization in a finite time is not guaranteed and, when it happens, it is controlled by the ``energy resources'' of the state. 

\smallskip

Previous example may be further generalized. Suppose that the spectral measure has form $\mu_u = A\delta_a + \nu_u$ for some $A > 0$ and $a \in \rr$, such that $\nu_u(\{a\}) = 0$ and $A > \nu_u(\sigma(H) - \{a\})$. Using
\begin{align*}
A & = |A e^{-iat}| = \left| \int_{\{a\}} e^{-i\omega t} \, \df\mu_u(\omega) \right| , \\
-\nu_u(\sigma(H) - \{a\}) & = -\int_{\sigma(H) - \{a\}} \df\mu_u(\omega) \le -\left| \int_{\sigma(H) - \{a\}} e^{-i\omega t} \, \df\mu_u(\omega) \right| ,
\end{align*}
and the triangle inequality, it follows that
\be
0 < A - \nu_u(\sigma(H) - \{a\}) \le \left| \int_{\{a\}} e^{-i\omega t} \, \df\mu_u(\omega) + \int_{\sigma(H) - \{a\}} e^{-i\omega t} \, \df\mu_u(\omega) \right| = |f_u(t)| ,
\ee
hence $t_\perp(u) = \infty$. In other words, if the spectral measure has a ``dominant atom'', the initial state never evolves to an orthogonal state. A converse of this result may be stated as follows: If there is a constant $C > 0$, such that $|f_u(t)| \ge C$ for all $t > 0$, then the Wiener's theorem \cite{RS3,Teschl} about the Ces{\`a}ro time average,
\be
C^2 \le \lim_{T \to \infty} \frac{1}{T} \, \int_0^T |\hat{\mu}_u(t)|^2 \, \df t = \sum_{x\in\rr} |\mu_u(\{x\})|^2 ,
\ee
implies that the measure $\mu_u$ must contain at least one atom.

\smallskip

We can construct examples in which function $f_u(t)$ has no zeros, orthogonalization time is $t_\perp(u) = \infty$, and $|f_u(t)|$ does not tend to zero as $t \to \infty$, nevertheless $\inf\{ |f_u(t)| : \, t > 0 \} = 0$. Take a Hamiltonian $H$ with at least three orthonormal eigenvectors $\{e_0,e_1,e_2\}$ and the corresponding eigenvalues $\{0,\omega_0,s\omega_0\}$, where $\omega_0 > 0$ and $s > 1$ is an irrational number. Linear combination
\be
u = \frac{1}{\sqrt{2}} \, e_0 + \frac{1}{2} \, (e_1 + e_2)
\ee
gives us
\be
f_u(t) = \frac{1}{4} \, \big( 2 + e^{-i\omega_0 t} + e^{-is\omega_0 t} \big) .
\ee
First, $f_u(t) = 0$ holds if and only if $e^{-i\omega_0 t} = e^{-is\omega_0 t} = -1$, that is $\omega_0 t = (2m+1)\pi$ and $s\omega_0 t = (2n+1)\pi$ for some $m,n \in \nn_0$. This implies $s = (2n+1)/(2m+1) \in \qq$, which is a contradiction. Next, note that
\be
|f_u(t)|^2 = \frac{1}{8} \, \big( 3 + 2\cos(\omega_0 t) + 2\cos(s \omega_0 t)  + \cos((s-1)\omega_0 t) \big) .
\ee
Therefore, using
\be
\lim_{T \to \infty} \frac{1}{T} \int_0^T \cos(\omega t) \, \df t = \lim_{T \to \infty} \frac{\sin(\omega T)}{\omega T} = 0
\ee
for $\omega \ne 0$, it follows that Ces{\`a}ro time average is nonzero
\be
\lim_{T \to \infty} \frac{1}{T} \int_0^T |f_u(t)|^2 \, \df t = \frac{3}{8} ,
\ee
which implies that $|f_u(t)|$ does not tend to zero. Namely, suppose that for any $\veps > 0$ there is $t_0 > 0$ such that $|f_u(t)| < \veps$ for all $t \ge t_0$. Then, using the abbreviation $M \defeq \max_{[0,t_0]} |f_u(t)|$, we have 
\be
\lim_{T \to \infty} \frac{1}{T} \int_0^T |f_u(t)|^2 \, \df t \le \lim_{T \to \infty} \Big( \frac{t_0 M^2}{T} + \big( 1 - (t_0/T) \big) \veps \Big) = \veps ,
\ee
which is in contradiction with the previous limit. Finally, by the continuity, for any $\veps > 0$ there is $\delta > 0$ such that $|\theta| < \delta$ implies $|1 - e^{-i\theta}| = 2|\sin(\theta/2)| < 4\veps$. Then, by the refinement of the Dirichlet's approximation theorem \cite{Scott40} (approximation of an irrational number by odd/odd rational number) we know that for any $\delta > 0$ there are $m,n \in \nn$ such that $|(2m+1)s - (2n+1)| < \delta$. So, if we choose $t = (2m+1)/\omega_0$ and $\theta = s\omega_0 t - (2n+1)$, we have simultaneously $|1 + e^{-i\omega_0 t}| = 0$ and 
\be
|1 + e^{-is\omega_0 t}| = |1 + e^{-(2n+1)i} e^{-i\theta}| = |1 - e^{-i\theta}| < 4\veps ,
\ee
which imply $|f_u(t)| < \veps$.

\smallskip

Even with the non-atomic spectral measures, orthogonalization time can be infinite. As an elementary example we may take Gaussian wave function
\be
\psi(x) = \left( \frac{2a}{\pi} \right)^{\!\frac{3}{4}} e^{-a|x|^2} \in L^2(\rr^3) ,
\ee
with $a > 0$, whose Fourier transform is
\be
\hat{\psi}(p) = \frac{1}{(2\pi)^{3/2}} \int_\rr e^{-ip\cdot x} \psi(x) \, \df^3 x = \frac{e^{-|p|^2/(4a)}}{(2\pi a)^{3/4}} .
\ee
If the dynamics is generated by the free particle Hamiltonian $H = -\Delta$, then the associated spectral measure is purely absolutely continuous (see Theorem 7.17 in \cite{Teschl})
\be
\df\mu_\psi(\omega) = 1_{[0,\infty)}(\omega) \left( \frac{\omega}{2\pi a^3} \right)^{\!\frac{1}{2}} \, e^{-\omega/(2a)} \, \df\omega ,
\ee
from where it follows that
\be
|f_\psi(t)|^2 = \frac{1}{(1 + 4a^2 t^2)^{3/2}}
\ee
does not have any zeros for $t > 0$, although $|f_\psi(t)| \to 0$ as $t \to \infty$.

\bigskip

\section{Quantum speed limits} 

Given that orthogonalization time is finite, we would like to understand how is it bounded by some given energy constraints. The strategy we shall use is based on auxiliary nonnegative functions which have only finite number (typically, two or three) of zeros. This idea has its roots in the work of Margoulis and Levitin \cite{ML98}, who introduced $\eta_{\mathrm{ML}} : [0,\infty) \to \rr$,
\be
\eta_{\mathrm{ML}}(x) = \cos x - 1 + 2 \, \frac{x}{\pi} + \frac{2}{\pi} \, \sin x ,
\ee
and subsequent, more systematic approach of Luo and Zhang\footnote{Zieli{\'n}ski and Zych \cite{ZZ06} use the same family of functions as Luo and Zhang.} \cite{LZ05}, with the family of functions $\eta_{\mathrm{LZ},p} : [0,\infty) \to \rr$ for $p > 0$,
\be
\eta_{\mathrm{LZ},p}(x) = \cos x - 1 + 2 \left( \frac{x}{\pi} \right)^p + \frac{2p}{\pi} \, \sin x .
\ee
Note that $\eta_{\mathrm{ML}} = \eta_{\mathrm{LZ},1}$. In their analysis of the MT bound, Kosi{\'n}ski and Zych \cite{KZ06} have introduced a family of functions $\eta_{\mathrm{KZ},\alpha} : \rr \to \rr$, for $\alpha \in \rr$,
\be
\eta_{\mathrm{KZ},\alpha}(x) = (x+\alpha)^2 - \frac{\pi^2}{4} + \pi\cos(x+\alpha) ,
\ee
while Levitin and Toffoli \cite{LT09} have utilized the function $\eta_{\mathrm{LT}} : \rr \to \rr$
\be
\eta_{\mathrm{LT}}(x) = \cos x - 1 + 2 \left( \frac{x}{\pi} \right)^2 + \frac{4}{\pi^2} \, x \sin x .
\ee
Unfortunately, most of the aforementioned papers \cite{ML98,KZ06,ZZ06,LT09} state the properties of these functions $\eta_*$ without proof, whereas Luo and Zhang \cite{LZ05} provide a complete, but somewhat cumbersome argument. To complement these existing results, we develop a systematic approach, presented in the Appendix (Theorems \ref{tm:etaKZ}, \ref{tm:etaLT} and \ref{tm:etaLZ}), that establishes the key properties of the functions $\eta_*$ through relatively concise proofs.

\smallskip

For any self-adjoint operator $A$, we define its and higher moments $\lb A^n \rb_u$ with respect to an unit vector $u \in D(A^n)$ for any $n \in \nn$,
\be
\lb A^n \rb_u \defeq \lb u, A^n u \rb .
\ee
In particular, operator's average value with respect to the unit vector $u \in D(A)$ is $\lb A \rb_u = \lb u, Au \rb$. For example, $n$-th energy moment may be written, using (\ref{eq:g(H)}), as an integral,
\be
\lb H^n \rb_u \defeq \lb u, H^n u \rb = \int_{\sigma(H)} \omega^n \, \df\mu_u(\omega) .
\ee
Furthermore, we define operator's standard deviation $\Delta_u A$ with respect to the unit vector $u \in D(A)$, as
\be
\Delta_u A \defeq \norm{(A - \lb A \rb_u I) u} .
\ee
Standard deviation vanishes, $\Delta_u A = 0$, if and only if $u$ is a unit eigenvector of $A$. It is important to note that $\Delta_u A$ is invariant with respect to the ``shift'' $A \to A + c I$ with a constant $c \in \rr$. Namely, any operator $A + cI$ is again self-adjoint, with equal domain $D(A + cI) = D(A)$, $\lb A +cI \rb_u = \lb A \rb_u + c$ and $\Delta_u (A + cI) = \Delta_u A$. If $u \in D(A^2)$, we may write
\be
(\Delta_u A)^2 = \lb A^2 \rb_u - \lb A \rb_u^2 .
\ee

\smallskip

In all proofs that follow, we rely on spectral representation of the real and imaginary parts of the function $f_u$,
\begin{align}
\Ree \, f_u(t) & = \int_{\sigma(H)} \cos(\omega t) \, \df\mu_u(\omega) , \\
\Imm \, f_u(t) & = -\int_{\sigma(H)} \sin(\omega t) \, \df\mu_u(\omega) .
\end{align}

\btm[Mandelstam--Tamm \cite{MT45}; \cite{KZ06,LT09}]\label{tm:MT}
Let $\HH$ be a complex separable Hilbert space, with $\dim\HH \ge 2$, and let $H : D(H) \to \HH$ be a densely defined, self-adjoint Hamiltonian. Then for each unit vector $u \in D(H^2)$
\be
t_\perp(u) \ge \frac{\pi}{2\Delta_u H} .
\ee
\etm

\noindent
\emph{Proof}. If the vector $u$ is an eigenvector of the Hamiltonian $H$, then the orthogonalization time is infinite, which is consistent with the statement of the theorem under the tacit convention that $1/\infty = 0$. We may therefore assume that $u$ is not an eigenvector of the Hamiltonian $H$, which implies that $\Delta_u H > 0$. We shall present two independent proofs.

\smallskip

(a) \emph{Kosi{\'n}ski--Zych approach}. Integral of the function $\eta_{\mathrm{KZ},\alpha}$ can be decomposed as
\begin{align*}
\int_{\sigma(H)} \eta_{\mathrm{KZ},\alpha}(\omega t) \, \df\mu_u(\omega) = t^2 \lb H^2 \rb_u & + 2\alpha t \lb H \rb_u + \alpha^2 - \frac{\pi^2}{4} + \nonumber\\
 & + \pi \big( \!\cos\alpha \, \Ree\, f_u(t) + i\sin\alpha \, \Imm\,f_u(t) \big) .
\end{align*}
In particular, at the orthogonalization time $t = t_\perp(u)$ we have $f_u(t_\perp(u)) = 0$, thus
\benn
t_\perp(u)^2 \lb H^2 \rb_u + 2\alpha t_\perp(u) \lb H \rb_u + \alpha^2 - \frac{\pi^2}{4} = \int_{\sigma(H)} \eta_{\mathrm{KZ},\alpha}(\omega t_\perp(u)) \, \df\mu_u(\omega) .
\eenn
Using the abbreviation
\be\label{eq:tauMT}
\tau_\mathrm{MT}(u) \defeq \frac{\pi}{2\Delta_u H} ,
\ee
we can conveniently rewrite last relation as
\be
(t_\perp(u)/\tau_\mathrm{MT}(u))^2 - 1 + \frac{4}{\pi^2} \, (\lb H \rb_u t_\perp(u) + \alpha)^2 = \frac{4}{\pi^2} \int_{\sigma(H)} \eta_{\mathrm{KZ},\alpha}(\omega t_\perp(u)) \, \df\mu_u(\omega) .
\ee
We know that the function $\eta_{\mathrm{KZ},\alpha}$ is nonnegative by Theorem \ref{tm:etaKZ}, thus the same holds for the left hand side for any choice of the parameter $\alpha$. One can always choose $\alpha = -\lb H \rb_u t_\perp(u)$, which immediately implies that $t_\perp(u) \ge \tau_\mathrm{MT}(u)$.

\smallskip

(b) \emph{Levitin--Toffoli approach}. Let us look at the spectral representation of the absolute value of $|f_u|$. Relying on the Fubini's theorem \cite{Axler,Teschl} we have
\begin{align*}
|f_u(t)|^2 & = \left( \int_{\sigma(H)} e^{-i\omega t} \, \df\mu_u(\omega) \right) \left( \int_{\sigma(H)} e^{+i\omega' t} \, \df\mu_u(\omega') \right) \nonumber\\
 & = \int_{\sigma(H)^2} e^{-i(\omega-\omega')t} \, \df(\mu_u \otimes \mu_u)(\omega,\omega') \nonumber\\
 & = \int_{\sigma(H)^2} \cos((\omega-\omega')t) \, \df(\mu_u \otimes \mu_u)(\omega,\omega') ,
\end{align*}
In the third line we have used the fact that $\sin((\omega-\omega')t)$ is odd with respect to $\omega \leftrightarrow \omega'$ exchange. Now, as
\benn
\int_{\sigma(H)^2} (\omega - \omega')^2 \, \df(\mu_u \otimes \mu_u)(\omega,\omega') = 2\big( \!\lb H^2 \rb_u - \lb  H \rb_u^2 \big) = 2 (\Delta_u H)^2 ,
\eenn
and, by the dominated convergence \cite{Axler},
\benn
\frac{\df}{\df t} \, |f_u(t)|^2 = -\int_{\sigma(H)^2} (\omega-\omega')\sin((\omega-\omega')t) \, \df(\mu_u \otimes \mu_u)(\omega,\omega') ,
\eenn
it follows that
\benn
\int_{\sigma(H)^2} \eta_\mathrm{LT}((\omega - \omega')t) \, \df(\mu_u \otimes \mu_u)(\omega,\omega') = |f_u(t)|^2 - 1 + \frac{4t^2}{\pi^2} \, (\Delta_u H)^2 - \frac{4t}{\pi^2} \, \frac{\df}{\df t} \, |f_u(t)|^2 .
\eenn
At the orthogonalization time we have $f_u(t_\perp(u)) = 0$ and $\df|f_u(t_\perp(u))|^2/\df t = 0$. With the abbreviation (\ref{eq:tauMT}) we can write the key relation as
\be\label{eq:ttauMT}
\left( \frac{t_\perp(u)}{\tau_\mathrm{MT}(u)} \right)^{\!2} = 1 + \int_{\sigma(H)^2} \eta_\mathrm{LT}((\omega - \omega') t_\perp(u)) \, \df(\mu_u \otimes \mu_u)(\omega,\omega') .
\ee
Finally, by the Theorem \ref{tm:etaLT} we know that the function $\eta_\mathrm{LT}$ is nonnegative, which implies that $t_\perp(u) \ge \tau_\mathrm{MT}(u)$. \qed

\bigskip

For the other bound we shall focus on Hamiltonians bounded from below, in the sense that there is $E_0 \in \rr$, such that $H - E_0 I \ge 0$ (recall that a densely defined operator $A$ is called nonnegative, written $A \ge 0$, if $\lb v, Av \rb \ge 0$ for all $v \in D(A)$). Note that $\sigma(H - E_0 I) \subseteq [0,\infty)$. Using (\ref{eq:funcalc}) we introduce a family of operators for $p > 0$,
\be
(H-E_0 I)^p = \int_{\sigma(H)} (\omega - E_0)^p \, \df P_H(\omega) ,
\ee
with the domains
\be
D((H - E_0 I)^p) = \left\{ u \in \HH : \int_{\sigma(H)} (\omega - E_0)^{2p} \, \df\mu_u(\omega) < \infty \right\} .
\ee
For any $p > 0$, the $p$-th energy moment of the ``shifted'' Hamiltonian is defined as
\be
\lb (H-E_0 I)^p \rb_u \defeq \int_{\sigma(H)} (\omega - E_0)^p \, \df\mu_u(\omega)
\ee
for any $u \in D((H - E_0 I)^p)$. As the function $h(\omega) = (\omega - E_0)^p$ is nonnegative on the domain $[E_0,\infty)$, with a single zero $\omega_0 = E_0$, Lemma \ref{lm:intXf} implies that $\lb (H-E_0 I)^p \rb_u = 0$ if and only if $E_0 \in \sigma(H)$ and $\mu_u = \delta_{E_0}$. Moreover, Lemma \ref{lm:umudelta} implies that in this case $u$ is in fact an eigenvector, $u \in \Ker(H - E_0 I)$.

\medskip

\btm[Margoulis--Levitin \cite{ML98}; Luo--Zhang \cite{LZ05}]\label{tm:MLLZ}
Let $\HH$ be a complex separable Hilbert space, with $\dim\HH \ge 2$, and let $H : D(H) \to \HH$ be a densely defined, self-adjoint Hamiltonian, bounded from below, such that $H - E_0 I \ge 0$ for $E_0 = \inf \sigma(H)$. Then for each $p > 0$ and unit vector $u \in D((H - E_0 I)^p)$ 
\be
t_\perp(u) \ge \frac{\pi}{(2\lb (H - E_0 I)^p \rb_u)^{1/p}} .
\ee
In particular, $p=1$ case is the Margoulis--Levitin bound \cite{ML98}.
\etm

\noindent
\emph{Proof}. Just as in the proof of the previous theorem, if the vector $u$ is an eigenvector of the Hamiltonian $H$, then the orthogonalization time is infinite, which is consistent with the statement of the theorem under the tacit convention that $1/\infty = 0$. We may therefore assume that $u$ is not an eigenvector of the Hamiltonian $H$, which implies that $\lb (H - E_0 I)^p \rb_u > 0$. Let us define a positive operator $K \defeq H - E_0 I$, with the domain $D(K) = D(H)$ and spectrum $\sigma(K) \subseteq [0,\infty)$. Accordingly, it is convenient to introduce shifted measure $\nu_u$ on $\rr$, such that $\nu_u(B) = \mu_u(B + E_0)$ for any Borel set $B \subseteq \rr$, and new function
\be
g_u(t) \defeq e^{+iE_0 t} f_u(t) = \int_{[0,\infty)} e^{-i\omega t} \, \df\nu_u(\omega) .
\ee
Then
\benn
\int_{[0,\infty)} \eta_{\mathrm{LZ},p}(\omega t) \, \df\nu_u(\omega) = \Ree \, g_u(t) - 1 + 2 (t/\pi)^p \lb K^p \rb_u - \frac{2p}{\pi} \, \Imm \, g_u(t) .
\eenn
At the orthogonalization time we have $g_u(t_\perp(u)) = 0$, hence
\benn
2\left( \frac{t_\perp(u)}{\pi} \right)^{\!p} \lb K^p \rb_u = 1 + \int_{[0,\infty)} \eta_{\mathrm{LZ},p}(\omega t_\perp(u)) \, \df\nu_u(\omega) .
\eenn
Using the abbreviation
\be
\tau_{\mathrm{LZ},p}(u) \defeq \frac{\pi}{(2\lb (H - E_0 I)^p \rb_u)^{1/p}}
\ee
we can conveniently rewrite last relation as
\be\label{eq:ttauLZ}
\left( \frac{t_\perp(u)}{\tau_{\mathrm{LZ},p}(u)} \right)^{\!p} = 1 + \int_{[0,\infty)} \eta_{\mathrm{LZ},p}(\omega t_\perp(u)) \, \df\nu_u(\omega)
\ee
Finally, by the Theorem \ref{tm:etaLT} we know that the function $\eta_{\mathrm{LZ},p}$ is nonnegative, which implies that $t_\perp(u) \ge \tau_{\mathrm{LZ},p}(u)$. \qed

\medskip

The particular choice of $E_0 = \inf \sigma(H)$, used in the theorem above, deserves a brief \emph{a posteriori} justification. Proof of the theorem could be repeated verbatim for any $E'_0 < E_0$, but as $\pi/(2\lb (H - E'_0 I)^p \rb_u)^{1/p} < \pi/(2\lb (H - E_0 I)^p \rb_u)^{1/p}$, resulting bounds would be strictly weaker. On the other hand, for any $E'_0 > E_0$ the proof fails because the support of the shifted measure $\nu_u$ intersects the negative real axis, where we no longer have control over behaviour of the auxiliary functions $\eta_{\mathrm{LZ},p}$.

\smallskip

For each probability measure $\mu$ on $[0,\infty)$ and real numbers $0 < p < q$ we know, via H\"older inequality \cite{AF,Folland}, that $L^q([0,\infty),\df\mu) \subseteq L^p([0,\infty),\df\mu)$. This result, in our context, immediately implies that $D(K^q) \subseteq D(K^p)$ (it is enough to note that $\omega^{2p} \le 1 + \omega^{2q}$ for all $\omega \ge 0$ and integrate with the measure $\nu_u$). So, if we define $p_*(u) = \sup \{p > 0: \, u \in D(K^p) \}$, then $u \in D(K^p)$ for each $0 < p < p_*(u)$. Given that $p_*(u) > 0$, one might try to identify the strongest LZ bound, which amounts to determining the supremum $\sup \{ \tau_{\mathrm{LZ},p}(u) : 0 < p < p_*(u) \}$. The key quantity for this question is the infimum $\inf\{ J_u(p) : \, 0 < p < p_*(u) \}$ of the function
\be
J_u(p) \defeq \left( 2 \int_{[0,\infty)} x^p \, \df\nu_u(x) \right)^{\!\frac{1}{p}} ,
\ee
with two competing factors, as $p < q$ implies $2^{1/p} > 2^{1/q}$ and $\norm{x}_p < \norm{x}_q$. Using the dominated convergence theorem, and relying on the elementary limit $\lim_{p\downarrow 0} x^p = 1$ for each $x > 0$, we have
\be
\lim_{p\downarrow 0} \int_{[0,\infty)} x^p \, \df\nu_u(x) = \mu((0,\infty)) .
\ee
If $0 \le \mu((0,\infty)) < 1/2$, limit $\lim_{p\downarrow 0} J(p) = 0$ implies $\lim_{p\downarrow 0} \tau_{\mathrm{LZ},p}(u) = \infty$. This is consistent with the discussion in Section 2, in which we have noticed that the presence of the ``dominant atom'' (here at $x=0$) implies infinite orthogonalization time. In the other cases, when $1/2 \le \mu((0,\infty)) \le 1$, the answer becomes more delicate. For example, if $\mu((0,\infty)) = 1/2$, we may introduced a rescaled probability measure $\rho_u$ on $(0,\infty)$, via $\rho_u(B) = 2\nu_u(B)$ for all Borel sets $B \subseteq (0,\infty)$, which allows us to identify $J_u(p) = \norm{x}_{L^p((0,\infty),\df\rho_u)}$. Again, norm inequalities $\norm{x}_p < \norm{x}_q$ for $p < q$, imply that $J_u$ is a nondecreasing function and (see e.g.~problem 8, chapter 6.1, in \cite{Folland})
\be
\lim_{p\downarrow 0} J(p) = \exp\left( \int_{(0,\infty)} \ln x \, \df\rho_u(x) \right) = \exp\left( 2\int_{(0,\infty)} \ln x \, \df\nu_u(x) \right) .
\ee
A more detailed characterization of the ordering of the LZ bounds in terms of finer properties of the spectral measure will not be pursued here.

\bigskip

\section{Saturation of the bounds} 

For each established bound $\tau_*(u)$ one might ask whether there exists a state $u_s$ that saturates the bound under the fixed energy resources relevant to it and, if so, to what extent that state is unique. Relying on lemmas proven in the Appendix B, we have a simple characterization of the associated spectral measure.

\btm\label{tm:satur}
Quantum speed bounds, proven in Theorems \ref{tm:MT} and \ref{tm:MLLZ}, are saturated if and only if
\be
\mu_{u_s} = \frac{1}{2} \, \big( \delta_{E_-} + \delta_{E_+} \big) ,
\ee
with $E_- < E_+$. In the case of ML and LZ bounds, $E_- = E_0 = \inf\sigma(H)$.
\etm

\noindent
\emph{Proof}. From the equation (\ref{eq:ttauMT}) we see that $t_\perp(u_s) = \tau_\mathrm{MT}(u_s)$ if and only if 
\be
\int_{\sigma(H)^2} \eta_\mathrm{LT}((\omega - \omega') t_\perp(u_s)) \, \df(\mu_{u_s} \otimes \mu_{u_s})(\omega,\omega') = 0 ,
\ee
while from the equation (\ref{eq:ttauLZ}) we see that $t_\perp(u_s) = \tau_{\mathrm{LZ},p}(u_s)$ if and only if 
\be
\int_{[0,\infty)} \eta_{\mathrm{LZ},p}(\omega t_\perp(u_s)) \, \df\nu_{u_s}(\omega) = 0 .
\ee
Then, using Theorem \ref{tm:etaLT} and Lemma \ref{lm:mumu}, that is Theorem \ref{tm:etaLZ} and Lemma \ref{lm:intXf}, it follows that $\mu_u = a\delta_{E_-} + (1-a)\delta_{E_+}$, with $0 \le a \le 1$ and $E_- \ne E_+$. Using equation (\ref{eq:fu}) we have
\benn
f_u(t) = a e^{-iE_- t} + (1-a) e^{-iE_+ t} .
\eenn
As was shown in the Section 2, this function leads to a finite orthogonalization time if and only if $a = 1-a$, that is $a = 1/2$. \qed

\medskip

The unresolved question in the theorem above is the exact form of the saturating vector $u_s$. In the nondegenerate case, when $\dim\Ker(H - E_- I) = 1$ and $\dim\Ker(H - E_+ I) = 1$, Lemma \ref{lm:umudelta} implies that
\be
u_s = \frac{1}{\sqrt{2}} \, \big( e_- + e_+ \big) ,
\ee
where $e_+$ and $e_-$ are orthonormal eigenvectors, $e_\pm \in \Ker(H - E_\pm I)$. However, if either $\dim\Ker(H - E_- I) \ge 2$ or $\dim\Ker(H - E_+ I) \ge 2$ (or both), we encounter an ambiguity. Namely, if $\{e_1,\dots,e_m\}$ are orthonormal vectors from $\Ker(H - E_- I)$, then all vectors $u = \sum_{i=1}^m c_i e_i$ with the coefficients $c_i \in \cc^\times$, normalized such that $\sum_{i=1}^m |c_i|^2 = 1$, share the same associated spectral measure, $\mu_u = \delta_{E_-}$ (and likewise for $E_+$). Fortunately, this ambiguity is not relevant for most of the energy computations considered in this paper.

\smallskip

Energy uncertainty of the saturating state is
\be
\Delta_{u_s} H = \frac{E_+ - E_-}{2} .
\ee
Furthermore, $u_s \in D((H - E_- I)^p)$ for all $p > 0$, and
\be
2\lb (H - E_- I)^p \rb_{u_s} = 2\int_{\sigma(H)} (\omega - E_-)^p \, \df\mu_{u_s}(\omega) = (E_+ - E_-)^p .
\ee
Given that $E_0 = \inf\sigma(H)$ is an eigenvalue of $H$, we may take $E_- = E_0$, with an immediate consequence that $\tau_\mathrm{MT}(u_s) = \tau_{\mathrm{LZ},p}(u_s) = \pi/(E_+ - E_0)$ for all $p > 0$. Orthogonalization time of the saturating state $u_s$ decreases as we take larger eigenvalue $E_+$ available in the spectrum of the Hamiltonian $H$.

\bigskip

\section{Near saturation of the bounds} 

Theorem \ref{tm:satur} reveals unique form of the spectral measure associated with the state which saturates MT and LZ family of QSL. Yet, this does not answers the question if there are other states whose orthogonalization time may come arbitrarily close to these bounds, or there is some kind of gap. More concretely, we would like to understand what kind of behaviour one might encounter with the states whose associated spectral measure is not pure point, as in the saturating case, but absolutely continuous or singular continuous with respect to the Lebesgue measure. A key insight is offered by the following theorem.

\btm\label{tm:near}
Let $f : [0,\infty) \to [0,\infty)$ be a continuous nonnegative function with a nonempty finite set of zeros $Z = f^{-1}(\{0\})$, and suppose that there exist $b,R > 0$ such that $f(x) > b$ for all $x > R$. Furthermore, let $\{ \mu_\veps : \, \veps > 0 \}$ be a family of positive Borel measures on $[0,\infty)$, such that $\mu_\veps([0,\infty)) = 1$ and
\be
\int_{[0,\infty)} f(x) \, \df\mu_\veps(x) = \veps .
\ee
Then $\lim_{\veps\downarrow 0} \mu_\veps(N_\delta) = 1$ for $N_\delta = \{ x \in [0,\infty) : \, d(x,Z) < \delta \}$ with fixed $\delta > 0$.
\etm

\noindent
\emph{Proof}. For each $\delta > 0$ the complement $N_\delta^c = [0,\infty) - N_\delta$ is a closed set (in the subspace topology) and the intersection $[0,R] \cap N_\delta^c$ is a compact set. Hence, by the Weierstra\ss~theorem, the continuous function $f$ on it attains a minimum $m_\delta > 0$. We can define $c_\delta \defeq \min \{b, m_\delta \} > 0$, so that $f(x) \ge c_\delta$ for each $x \in N_\delta^c$. This gives us an inequality
\be
\veps \ge \int_{N_\delta^c} f(x) \, \df\mu_\veps(x) \ge c_\delta \, \mu_\veps(N_\delta^c) ,
\ee
from where the claim immediately follows. \qed

\medskip

Suppose that we have a family of vectors $u_\veps \in \HH$ for $\veps > 0$, such that
\be
\int_{\sigma(H)} \eta_*(x) \, \df\mu_{u_\veps}(x) = \veps ,
\ee
for some auxiliary nonegative function $\eta_*$. Then $t_\perp(u)/\tau_*(u) = 1 + \veps'$, where $\veps' = O(\veps)$ as $\veps\downarrow 0$ (precise relation between $\veps$ and $\veps'$ depend on particular QSL, but it is not important for this discussion). In other words, such states nearly saturate the pertaining QSL. Now, all auxiliary functions $\eta_*$, introduced in Section 3, satisfy assumptions from the Theorem \ref{tm:near}: they are nonnegative, have nonempty finite set of zeros and, since they grow as $O(x^p)$ for some power $p>0$ as $x \to \infty$, there are $b,R > 0$ such that $\eta_*(x) > b$ for all $x > R$. Therefore, the theorem is giving us an insight that the spectral measure is ``concentrated around zeros of $\eta_*$'', in the sense that $\lim_{\veps\downarrow 0} \mu_\veps(N_\delta) = 1$ for every fixed $\delta$. We shall construct and investigate two explicit examples with the purely absolutely continuous and purely singular continuous spectral measures, such that quantum speed limits are ``nearly saturated''. In both cases we shall, for simplicity, assume that the self-adjoint Hamiltonian $H$ is bounded from below and $E_0 \defeq \inf\sigma(H)$.

\medskip

\noindent
\textbf{5.1. Absolutely continuous case}. Let $\rho : \rr \to [0,\infty)$ be a nonnegative Borel function, with $\supp\rho \subseteq [0,1]$, normalized such that
\be
\int_0^1 \rho(x) \, \df x = 1 .
\ee
For each $\veps > 0$ we define a rescaled function $\rho_\veps : \rr \to [0,\infty)$ via $\rho_\veps(x) = \rho(x/\veps)/\veps$, whose support is $\supp\rho_\veps \subseteq [0,\veps]$. The associated moments
\be
M_n(\rho) \defeq \int_0^1 x^n \rho(x) \, \df x
\ee
satisfy inequalities $M_n(\rho)^m \le M_m(\rho)^n$ for all integers $1 \le n \le m$. This is a direct consequence of the inequalities $\norm{f}_n \le \norm{f}_m$ on spaces $L^p([0,1], \rho(x)\,\df x)$ with the probability measure $\df\mu(x) = \rho(x)\,\df x$ (see e.g.~Proposition 6.12 in \cite{Folland}). We introduce a family of measures
\be
\df\mu_{u_\veps}(\omega) = \frac{1}{2} \, \big( \rho_\veps(\omega - E_0) + \rho_\veps(\omega - E') \big) \, \df\omega ,
\ee
with the energy $E' > E_0$, which will be determined such that the bound $\tau_*(u_\veps)$ is equal to the pertaining $\tau_*(u_s) = \pi/(E_+ - E_0)$ with some given energy $E_+ > E_0$. Also, for consistency, we demand that $\supp\mu_{u_\veps} \subseteq \sigma_{\mathrm{ac}}(H)$. Last condition is immediately satisfied with e.g.~Schr\"odinger operator $H = -\Delta + E_0 I$, which has $\sigma(H) = \sigma_{\mathrm{ac}}(H) = [E_0,\infty)$ (see Theorem 7.17 in \cite{Teschl}).

\smallskip

Let us look more closely into the energy quantities. For the MT bound we need energy uncertainty,
\be\label{eq:acDelta}
\big( \Delta_{u_\veps} (H - E_0 I) \big)^2 = \veps^2 \big( M_2(\rho) - M_1(\rho)^2 \big) + \frac{1}{4} \, (E' - E_0)^2 , 
\ee
and for the ML and general LZ bounds we need energy moments (for simplicity we shall focus only on the integer moments, $n\in\nn$)
\be\label{eq:acaven}
2\lb (H - E_0 I)^n \rb_{u_\veps} = \veps^n M_n(\rho) + \sum_{k=0}^n \binom{n}{k} \veps^k M_k(\rho) (E' - E_0)^{n-k} .
\ee
While, in principle, we could solve conditions 
\benn
2\Delta_{u_\veps} (H - E_0 I) = E_+ - E_0 \quad \textrm{or} \quad 2\lb (H - E_0 I)^n \rb_{u_\veps} = (E_+ - E_0)^n
\eenn
for $E'$, the crucial information is that in all these cases solution satisfies $E' > E_+$ and $E' - E_+ = O(\veps)$ as $\veps\downarrow 0$.

\smallskip

A convenient way to write the key function $f_{u_\veps}$ in this case is
\begin{align}
f_{u_\veps}(t) & = \frac{1}{2} \int_{\sigma(H)} e^{-i\omega t} \big( \rho_\veps(\omega - E_0) + \rho_\veps(\omega - E') \big) \, \df\omega \nonumber\\
 & = \frac{\sqrt{2\pi}}{2} \, e^{-iE_0 t} \big( 1 + e^{-i(E' - E_0)t} \big) \hat{\rho}_\veps(t) ,
\end{align}
where $\hat{\rho}_\veps$ is the Fourier transform of $\rho_\veps$. Given that $\hat{\rho}_\veps(t) \ne 0$ on some interval $0 \le t < t_0$, the ortogonalization time is given by
\be
t_\perp(u_\veps) = \frac{\pi}{E' - E_0} ,
\ee
given that $t_\perp(u_\veps) < t_0$. Due to previous remark, we have $t_\perp(u_\veps) > \pi/(E_+ - E_0)$, so it does not saturate any of the QSL. However, $\lim_{\veps\downarrow 0} t_\perp(u_\veps) = \pi/(E_+ - E_0)$, so in this sense this is family of ``near-saturating'' states.

\smallskip

Technical condition on Fourier transform $\hat{\rho}_\veps(t)$ is not too strong, which can be illustrated as follows. We can, for example, choose $\rho(x) = 1_{[0,1]}(x)$, for which $\rho_\veps(x) = 1_{[0,\veps]}(x)/\veps$ and
\be
\hat{\rho}_\veps(t) = \frac{1}{\sqrt{2\pi}} \, \frac{1}{\veps} \int_0^\veps e^{-i\omega t} \, \df \omega = \frac{1}{\sqrt{2\pi}} \, \frac{\sin(\veps t/2)}{\veps t/2} \, e^{-i\veps t/2} .
\ee
Function $\hat{\rho}_\veps$ has no zeros on the interval $(0,2\pi/\veps)$. Therefore, condition $t_\perp(u_\veps) < t_0$ holds given that $\veps < 2(E'-E_0)$.

\medskip

\noindent
\textbf{5.2. Singular continuous case}. Construction of singular continuous examples will be analogous to the previous analysis, building upon the family of Cantor measures $\kappa_\theta$, supported on the (generalized) Cantor set $C_\theta$ with the parameter $0 < \theta < 1/2$. A brief overview to these measures is given in the Appendix C. Here we introduce a family of spectral measures for each $\veps > 0$,
\be
\mu_{u_\veps} = \frac{1}{2} \, \big( F_{\veps,*} \kappa_\theta + G_{\veps,*} \kappa_\theta \big)
\ee
with $F_\veps(x) = \veps x + E_0$ and $G_\veps(x) = \veps x + E'$ (functions $F_\veps$ and $G_\veps$ serve to push-forward measure $\kappa_\theta$ from its support $\Can_\theta \subseteq [0,1]$ to, respectively, $\veps\Can_\theta + E_0 \subseteq [E_0, E_0+\veps]$ and $\veps\Can_\theta + E' \subseteq [E', E'+\veps]$). As in the absolutely continuous case, the energy $E' > E_0$ will be determined such that the bound $\tau_*(u_\veps)$ agrees with the pertaining $\tau_*(u_s) = \pi/(E_+ - E_-)$, and for consistency we demand $\supp\mu_{u_\veps} \subseteq \sigma_{\mathrm{sc}}(H)$.

\smallskip

Canonical example of a Hamiltonian adequate for this case is the multiplication operator, $(H\psi)(x) = x\psi(x)$, on the Hilbert space $L^2([0,1],\df\kappa_\theta)$. Its spectrum is equal to the essential range of the function $\varphi(x) = x$ (see e.g.~Theorem 4.5 in \cite{Borthwick}), $\sigma(H) = C_\theta$. The associated projection-valued measure $P_H$ acts as a simple multiplication with the characteristic function, $(P_H(B)\psi)(x) = 1_B(x) \psi(x)$, for every Borel set $B \subseteq [0,1]$. Furthermore, for each nonzero $u \in L^2([0,1],\df\kappa_\theta)$, the associated spectral measure is purely singular continuous, as $\mu_u(B) = \lb u, 1_B u \rb = \kappa_\theta(1_B |u|^2)$, hence $\sigma(H) = \sigma_{\mathrm{sc}}(H)$ (it is far less trivial to find Schr\"odinger operator $H = -\Delta + V$ with such a spectrum \cite{Pearson78}).

\smallskip

Energy quantities may be expressed with the help of the moments of the measure $\kappa_\theta$ (\ref{eq:momkappa}),
\begin{align*}
\big( \Delta_{u_\veps} (H - E_0 I) \big)^2 & = \veps^2 \big( M_2(\kappa_\theta) - M_1(\kappa_\theta)^2 \big) + \frac{1}{4} \, (E' - E_0)^2 , \\
2\lb (H - E_0 I)^n \rb_{u_\veps} & = \int_{[E_0,E_0+\veps]} (\omega - E_0)^n \, \df(F_{\veps,*} \kappa_\theta)(\omega) + \int_{[E',E'+\veps]} (\omega - E_0)^n \, \df(G_{\veps,*} \kappa_\theta)(\omega) \\
 & = \int_{[0,1]} \veps^n \omega^n \, \df\kappa_\theta(\omega) + \int_{[0,1]} (\veps\omega + E' - E_0)^n \, \df\kappa_\theta(\omega) \\
 & = \veps^n M_n(\kappa_\theta) + \sum_{k=0}^n \binom{n}{k} \veps^k M_k(\kappa_\theta) (E' - E_0)^{n-k} .
\end{align*}
Structure of these expressions is identical to the absolute continuous ones (\ref{eq:acDelta})--(\ref{eq:acaven}), so that the solutions of constrains $2\Delta_{u_\veps} (H - E_0 I) = E_+ - E_0$ or $2\lb (H - E_0 I)^n \rb_{u_\veps} = (E_+ - E_0)^n$ again satisfies $E' > E_+$ and $E' - E_+ = O(\veps)$ as $\veps\downarrow 0$. Using results from the Appendix C, the key function $f_{u_\veps}$ may be written in the form
\begin{align}
f_{u_\veps}(t) & = \frac{\sqrt{2\pi}}{2} \, \big( 1 + e^{-i(E' - E_0)t} \big) e^{-iE_0 t} \hat{\kappa}_\theta(\veps t) \nonumber\\
 & = \frac{1}{2} \, \big( 1 + e^{-i(E' - E_0)t} \big) e^{-i(E_0 + \veps/2)t} \prod_{m=0}^\infty \cos((1-\theta)\theta^m \veps t/2) .
\end{align}
Product of the cosine functions above has no zeros on the interval $(0,t_0)$ for $t_0 = \pi/((1-\theta)\veps)$. Hence, assuming that $\veps < (E'-E_0)/(1-\theta)$, the orthogonalization time is defined by the condition $1 + e^{-i(E' - E_0)t_\perp(u_\veps)} = 0$, which gives $t_\perp(u_\veps) = \pi/(E' - E_0) > \pi/(E_+ - E_0)$ and $\lim_{\veps\downarrow 0} t_\perp(u_\veps) = \pi/(E_+ - E_0)$.

\bigskip

\section{Numerical range perspective} 

In order to complement previous analysis, we shall investigate another approach to QSLs, with a valuable geometric insight. First we define the limited energy resources with the compact set $S \subseteq \sigma(H)$, which has at least two elements. The associated spectral projection $P_H(S)$ defines the closed subspace $\HH_S \defeq P_H(S) \HH$. For any unit vector $u \in \HH_S$ we have $\mu_u(S) = \lb u, P_H(S)u \rb = \lb u, u \rb = 1$ and its average energy is
\be
\lb H \rb_u = \lb u, Hu \rb = \lb u, H P_H(S) u \rb = \int_{\sigma(H)} \omega 1_S(\omega) \, \df\mu_u(\omega) = \int_S \omega \, \df\mu_u(\omega) .
\ee
If we denote by $\omega_- = \inf S$ and $\omega_+ = \sup S$, energy bounds are revealed by the inequalities
\be
\omega_- = \int_S \omega_- \, \df\mu_u(\omega) \le \lb H \rb_u \le \int_S \omega_+ \, \df\mu_u(\omega) = \omega_+ .
\ee
Next, numerical range of an operator $A : D(A) \to \HH$ is defined as the set
\be
W(A) \defeq \{ \lb v, Av \rb : \, v \in D(A) , \, \norm{v} = 1 \} \subseteq \cc .
\ee
We are especially interested in the function
\be
W_S(t) \defeq W(e^{-iHt}|_{\HH_S}) ,
\ee
which maps real numbers (time $t$) to subsets of the complex plane. Now we can rephrase the ortogonalization problem as follows (this approach was briefly explored in \cite{VHS21}): Given a fixed time $t > 0$, is there any initial unit vector $u \in \HH_S$, such that $0 \in W_S(t)$? From this point of view, focus shifts to the question about the geometry of the numerical range.

\smallskip

Toeplitz--Hausdorff theorem states that the numerical range $W(A)$ for any operator $A$ is a convex subset of $\cc$ (a particularly elegant and concise proof, together with references to earlier proofs, may be found in \cite{Gustafson70}). Moreover, numerical range and spectrum of a normal operator $N$ are related by $\cl{W(N)} = \mathrm{ch}(\sigma(N))$, where $\mathrm{ch}$ denotes a convex hull of a set (see chapter 22 in \cite{Halmos}). In order to utilize this geometric connection, we shall need a refinement of the spectral mapping theorem \cite{Schm,Teschl}. We claim that
\be
\sigma(e^{-iHt}|_{\HH_S}) \subseteq \mathscr{A} \defeq \cl{ \{\exp(-i\omega t) : \, \omega \in S \} } .
\ee
For each $z \in \cc - \mathscr{A}$ the function $b_z : S \to \cc$, $b_z(\omega) = (e^{-i\omega t} - z)^{-1}$, is bounded by $|b_z(\omega)| \le 1/\delta$ for all $\omega \in S$, where $\delta \defeq d(z,\mathscr{A}) > 0$. This allows us to define the operator
\be
R_z \defeq \int_S b_z(\omega) \, \df P_H(\omega)
\ee
on the whole Hilbert space $\HH$. Namely, as $|b_z|^2$ is a bounded function,
\benn
D(R_z) = \left\{ v \in \HH : \int_S |b_z(\omega)|^2 \, \df\mu_v(\omega) < \infty \right\} = \HH .
\eenn
For each unit vector $u \in \HH$ we have (see Proposition 4.15 in \cite{Schm})
\be
\norm{R_z u}^2 = \int_S |b_z(\omega)|^2 \, \df\mu_u(\omega) \le \delta^{-2} \mu_u(S) = \delta^{-2} \norm{P_H(S) u}^2 \le \delta^{-2} \norm{u}^2 = \delta^{-2} ,
\ee
thus $R_z$ is a bounded operator. Also,
\be
P_H(S) R_z = \bigg( \int_S 1_S(\omega) \, \df P_H(\omega) \bigg) \bigg( \int_S b_z(\omega) \, \df P_H(\omega) \bigg) = \int_S b_z(\omega) \, \df P_H(\omega) = R_z
\ee
implies that $\Ran\,R_z \subseteq \HH_S$. Since our focus is on the operator $e^{-iHt}|_{\HH_S}$, with the range\footnote{This is a direct consequence of the commutator $P_H(S) e^{-iHt} = e^{-iHt} P_H(S)$, which holds for any $t \in \rr$.} $\Ran(e^{-iHt}|_{\HH_S}) = \HH_S$, corresponding candidate resolvent operator is the restriction $R_z|_{\HH_S}$. Functional calculus confirms that $R_z|_{\HH_S}$ is indeed a resolvent (see Proposition 4.12 in \cite{Schm}),
\be
R_z|_{\HH_S} (e^{-iHt} - zI)|_{\HH_S} = (e^{-iHt} - zI)|_{\HH_S} R_z|_{\HH_S} = \int_S 1_S(\omega) \, \df P_H(\omega) \, \Big|_{\HH_S} = I|_{\HH_S} .
\ee
Hence, $\cc - \mathscr{A} \subseteq \cc - \sigma(e^{-iHt}|_{\HH_S})$, which proves the claim. 

\smallskip

An elementary fact is that for any two nonempty subsets $A,B \subseteq \rr^n$, $A \subseteq B$ implies $\mathrm{ch}(A) \subseteq \mathrm{ch}(B)$ and $\cl{A} \subseteq \cl{B}$. Therefore,
\be
\cl{W_S(t)} = \mathrm{ch}\big(\sigma(e^{-iHt}|_{\HH_S})\big) \subseteq \mathrm{ch}\!\left( \cl{ \{\exp(-i\omega t) : \, \omega \in S \} } \right) ,
\ee
and, as by assumption $S \subseteq [\omega_-,\omega_+]$,
\be
\cl{W_S(t)} \subseteq \mathrm{ch}(\mathscr{C}) \qqd \mathscr{C} \defeq \{\exp(-i\omega t) : \, \omega \in [\omega_-,\omega_+] \} .
\ee
The set $\mathscr{C}$ is the arc of the unit circle in the complex plane, centered at the origin, with endpoints $e^{-i\omega_- t}$ and $e^{-i\omega_+ t}$, while its convex hull $\mathrm{ch}(\mathscr{C})$ is the corresponding circular segment, which may be minor, major, or the entire unit disc, depending on the phase difference $(\omega_+ - \omega_-)t$.

\smallskip

For any time $0 < t < \pi/(\omega_+ - \omega_-)$, the length of the arc $\mathscr{C}$ is smaller than $\pi$, which implies that $0 \notin \mathrm{ch}(\mathscr{C})$ and $0 \notin \cl{W_S(t)}$. In other words, for any unit vector $u \in \HH_S$ the orthogonalization time is bounded by $t_\perp(u) \ge \pi/(\omega_+ - \omega_-)$. At the critical moment $t_* = \pi/(\omega_+ - \omega_-)$ we have $0 \in \cl{W_S(t_*)}$, but this is not enough to conclude whether $0 \in W_S(t_*)$.

\bigskip

\section{Final remarks} 

We have revisited several bounds on orthogonalization time and generalized their proofs to the functional-analytic level of rigor in Theorems \ref{tm:MT} and \ref{tm:MLLZ}. As shown by the Theorem \ref{tm:satur}, these bounds are sharp and saturated for the states with the associated spectral measure which is a sum of two Dirac measures. Furthermore, spectral measures of near-saturating states were constrained by Theorem \ref{tm:near} and we have constructed examples with purely absolutely continuous and purely singular continuous spectral measures, whose orthogonalization time may come arbitrarily close to the bounds defined by the QSLs. Finally, in order to enrich our intuition about the evolution of the states, we have briefly analysed orthogonalization time with the operator's numerical range.

\smallskip

Nevertheless, many questions remain open and warrant further investigation. First of all, it is not clear if all auxiliary functions, that served as a backbone of the proofs in Sections 3 and 4, are just special cases of some larger class of related functions and if these can be used to gain  additional physical insights. A possible candidate is a family of functions
\benn
\eta_{p,q}(x) = \cos x - 1 + 2(x/\pi)^p + \frac{2p}{\pi} \, (x/\pi)^q \sin x ,
\eenn
which subsumes $\eta_{\mathrm{LZ,p}}$ for $q = 0$ and $\eta_\mathrm{LT}$ for $(p,q) = (2,1)$. We conjecture that $\eta_{p,q}$ satisfies the key properties, nonegativity and finite number of zeros, for a certain range of parameter values $(p,q)$. Also, it would be interesting to see if the approach based on the operator's numerical range and the accompanying geometric picture, could be refined for the various QSL bounds.

\smallskip

A more challenging objective lies in the rigorous generalizations of QSLs to mixed states \cite{Pfeifer93,PF95,CPBM18} and open systems \cite{TEDMF13,dCEPH13,DL13}. An important step in this direction has recently been taken by Becker and Datta \cite{BD20}, who established energy-constrained continuity bounds for quantum dynamical semigroups in infinite-dimensional systems, encompassing both unitary and open-system dynamics with potentially unbounded generators. Further progress in this direction may involve extending the framework to more general dynamical settings and establishing sharper QSLs under physically motivated energy constraints.

\bigskip

\section*{Acknowledgements}

The research was supported by the Croatian Science Foundation Project No.~IP-2025-02-8625, \emph{Quantum aspects of gravity}, by the European Union --- NextGenerationEU through the National Recovery and Resilience Plan 2021-2026, and by the project \emph{Implementation of cutting-edge research and its application as part of the Scientific Center of Excellence for Quantum and Complex Systems, and Representations of Lie Algebras}, Grant No. PK.1.1.10.0004, co-financed by the European Union through the European Regional Development Fund --- Competitiveness and Cohesion Programme 2021--2027.

\bigskip

\appendix 

\section{Auxiliary inequalities} 

In this Appendix we shall investigate key properties of the auxiliary functions $\eta_{\mathrm{KZ},\alpha}$, $\eta_\mathrm{LT}$ and $\eta_{\mathrm{LZ},p}$, introduced in Section 3. Most importantly, we shall prove that these are nonnegative functions on their respective domains, and each of them has a finite set of zeros. We begin with several elementary inequalities for trigonometric functions.

\smallskip

\blm\label{lm:trig}

\begin{align*}
x > |\!\sin x| , & \quad x > 0 \\
x^3/6 > x - \sin x , & \quad x > 0 \\
\tan x > x , & \quad 0 < x < \pi/2 \\
x \cot(x/2) > 2(1 - x/\pi) , & \quad 0 < x < \pi \\
-x \cot(x/2) > 2(x/\pi - 1) , & \quad \pi < x < 2\pi \\
\pi (x/\pi)^p + p\sin x > 0 , & \quad x > 0, \, p > 0
\end{align*}
\elm

\noindent
\emph{Proof}. The proof strategy rests upon a careful choice of functions with a convenient monotonicity and convexity properties. First, for $f(x) = x - \sin x$, we have $f'(x) = 1 - \cos x > 0$ almost everywhere, thus
\benn
f(x) = \int_0^x f'(y) \, \df y > 0
\eenn
for all $x > 0$. Same reasoning with the auxiliary functions, $g_1(x) = x^2 - \sin^2 x$, $g_2(x) = x^3/6 - x + \sin x$ and $g_3(x) = \tan x - x$, whose derivatives may be written in a manifestly nonnegative form, $g'_1(x) = 2x - \sin(2x)$, $g'_2(x) = 2((x/2)^2 - \sin^2(x/2))$ and $g'_3(x) = \tan^2 x$, proves the first three inequalities. Furthermore, function $h : [0,2\pi) \to \rr$,
\benn
h(x) = \left\{ \begin{array}{rl} 2 \, , & x = 0 \\ x\cot(x/2) \, , & 0 < x < 2\pi \end{array} \right.
\eenn
is concave on the interval $[0,\pi]$, hence the chord $\ell(x) = 2(1 - x/\pi)$ satisfies $\ell(x) < h(x)$ for all $0 < x < \pi$, which proves the fourth inequality. Moreover, $-h'(x) = (x-\sin x)/(1-\cos x) \ge x/2 > \pi/2 > 2/\pi$ for $\pi < x < 2\pi$ implies the fifth inequality.

Final inequality deserves a more careful treatment. Let $k_p : (0,\infty) \to \rr$, $k_p(x) = \pi(x/\pi)^p + p\sin x$. Due to periodicity of the trigonometric functions and monotonicity of $x \mapsto x^p$ for $p > 0$, we have 
\benn
k_p(x+2\pi) = k_p(x) + \pi^{1-p} \big( (x+2\pi)^p - x^p \big) > k_p(x)
\eenn
for all $x > 0$. Hence, it is enough to prove that $k_p(x) > 0$ for all $0 < x < 2\pi$. Inequality immediately follows for $0 < x \le \pi$, where $\sin x \ge 0$. For $\pi < x < 2\pi$ we have $\sin x = -\sin(x - \pi) > -(x-\pi)$, so that $k_p(x) > \pi((x/\pi)^p - p((x/\pi) - 1))$. If $0 < p \le 1$, then $(x/\pi)^p > 1 \ge p > p((x/\pi) - 1)$. If $p \ge 1$, using Bernoulli inequality, we have $(x/\pi)^p > 1 + p((x/\pi) - 1) > p((x/\pi) - 1)$, which proves the claim. \qed

\bigskip

Building on these inequalities the following three theorems\footnote{Proof strategies in Theorems \ref{tm:etaLT} and \ref{tm:etaLZ} originate from the the online discussion at the StackExchange forum, http://math.stackexchange.com/q/3952811, about the nonegativity of the function $\eta_\mathrm{ML}$.} establish key facts used in proofs of QSLs.

\btm\label{tm:etaKZ}
$\eta_{\mathrm{KZ},\alpha}(x) \ge 0$ for all $x,\alpha \in \rr$ and the equality holds if and only if $x \in \{-\alpha - \pi/2, -\alpha + \pi/2\}$. 
\etm

\noindent
\emph{Proof}. It is enough to proof the claim for
\benn
g(x) \defeq \eta_{\mathrm{KZ},0}(x) = x^2 - \frac{\pi^2}{4} + \pi\cos x ,
\eenn
from where the general theorem follows by translation. Function $g$ is even and
\benn
g(x+\pi) = g(x) + 2\pi \left( x + \frac{\pi}{2}  - \sin(x + \pi/2) \right) > g(x)
\eenn
for all $x \ge 0$. Also, we have $g(0) = \pi(4-\pi)/4 > 0$, $g(\pi/2) = 0$ and $g(\pi) = \pi(3\pi - 4)/4 > 0$. Hence, it is enough to prove that $g(x) > 0$ for $x \in (0,\pi/2) \cup (\pi/2,\pi)$. First, $g'(x) = 2x - \pi\sin x$ is a strictly convex function on the interval $(0,\pi/2)$, with $g'(0) = g'(\pi/2) = 0$, thus $g'(x) < 0$ for all $x \in (0,\pi/2)$. Moreover, $g'(x) = \pi(2x/\pi - \sin x) > \pi(1 - \sin x) > 0$ for all $x \in (\pi/2,\pi)$, which completes the proof. \qed

\medskip

\btm\label{tm:etaLT}
$\eta_\mathrm{LT}(x) \ge 0$ for all $x \in \rr$ and the equality holds if and only if $x \in \{-\pi,0,\pi\}$. 
\etm

\noindent
\emph{Proof}. It is easy to check that $\eta_\mathrm{LT}(x) = 0$ for $x \in \{-\pi,0,\pi\}$. Furthermore, $\eta_\mathrm{LT}$ is an even function and, due to periodicity of the trigonometric functions and Lemma \ref{lm:trig}, we have
\benn
\eta_\mathrm{LT}(x+2\pi) = \eta_\mathrm{LT}(x) + 8 \left( 1 + \frac{x+\sin x}{\pi} \right) > \eta_\mathrm{LT}(x)
\eenn
for all $x > 0$. Thus, it is enough to prove that $\eta_\mathrm{LT}(x) > 0$ for all $x \in (0,\pi) \cup (\pi,2\pi)$. We shall introduce an auxiliary function $G : (0,2\pi) \to \rr$,
\be
G(x) \defeq \frac{1-\cos x}{2(x/\pi)^2 + (4/\pi^2) x\sin x} ,
\ee
which satisfies $\lim_{\veps\downarrow 0} G(\veps) = \pi^2/12 \approx 0.822$, $G(\pi) = 1$ and $\lim_{\veps\downarrow 0} G(2\pi-\veps) = 0$. We claim that its derivative,
\be
G'(x) = \frac{2\pi^2 \sin x}{x^2} \, \frac{(x/2)^2 - \sin^2(x/2)}{(x + 2\sin x)^2} ,
\ee
satisfies $G'(x) > 0$ for $0 < x < \pi$ and $G'(x) < 0$ for $\pi < x < 2\pi$. Note that the expression above is well-defined for $x > 0$ as $x + 2\sin x > 0$ (sine function is positive for $0 < x < \pi$ and $x + 2\sin x \ge x - 2 > \pi - 2 > 0$ for $x\ge\pi$). Finally, by Lemma \ref{lm:trig}, $(x/2)^2 > \sin^2(x/2)$ for all $x > 0$, hence sign of $G'(x)$ is equal to sign of $\sin x$. This proves that $0 < G(x) < 1$ for all $x \in (0,\pi) \cup (\pi,2\pi)$, from where the main claim immediately follows. \qed

\medskip

\btm\label{tm:etaLZ}
$\eta_{\mathrm{LZ},p}(x) \ge 0$ for all $x \ge 0$ and the equality holds if and only if $x \in \{0,\pi\}$. 
\etm

\noindent
\emph{Proof}. It is easy to check that $\eta_{\mathrm{LZ},p}(x) = 0$ for $x \in \{0,\pi\}$. Moreover, due to periodicity of the trigonometric functions and monotonicity of $x \mapsto x^p$ for $p > 0$, we have
\benn
\eta_{\mathrm{LZ},p}(x+2\pi) = \eta_{\mathrm{LZ},p}(x) + \frac{2((x+2\pi)^p - x^p)}{\pi^p} > \eta_{\mathrm{LZ},p}(x)
\eenn
for all $x > 0$. Thus, it is enough to prove that $\eta_{\mathrm{LZ},p}(x) > 0$ for all $x \in (0,\pi) \cup (\pi,2\pi)$. We shall introduce a family of auxiliary functions $F_p : (0,2\pi) \to \rr$ for $p > 0$,
\be
F_p(x) \defeq \frac{1-\cos x}{2(x/\pi)^p + (2p/\pi)\sin x} ,
\ee
which satisfy $\lim_{\veps\downarrow 0} F_p(\veps) = 0$, $F(\pi) = 1$ and $\lim_{\veps\downarrow 0} F(2\pi-\veps) = 0$. We claim that its derivative,
\benn
F'(x) = \frac{\pi (x/\pi)^{p-1}}{2(\pi (x/\pi)^p + p\sin x)^2} \, \Big( p(1-\cos x) \big( (\pi/x)^{p-1} - 1 \big) + x\sin x \Big) ,
\eenn
satisfies $F'_p(x) > 0$ for $0 < x < \pi$ and $F'_p(x) < 0$ for $\pi < x < 2\pi$. Note that the expression above is well-defined for since by Lemma \ref{lm:trig} $\pi x^p + p\pi^p \sin x > 0$ holds for all $x > 0$. We shall split the proof into two subcases:

\begin{itemize}
\item[(a)] $p \ge 1$. For $0 < x < \pi$ we have $(\pi/x)^{p-1} \ge 1$ and $\sin x > 0$, thus $F'(x) > 0$. For $\pi < x < 2\pi$ we have $(\pi/x)^{p-1} \le 1$ and $\sin x < 0$, thus $F'(x) < 0$.

\item[(b)] $0 < p < 1$. Using inequalities from Lemma \ref{lm:trig} we have
\benn
\frac{x\sin x}{1 - \cos x} = x\cot(x/2) > 2(1 - x/\pi) > p(1 - x/\pi) > p(1 - (x/\pi)^{1-p})
\eenn
for all $0 < x < \pi$, hence $F'_p(x) > 0$, while
\benn
-\frac{x\sin x}{1 - \cos x} = -x \cot(x/2) > 2(x/\pi - 1) > p(x/\pi - 1) > p((\pi/x)^{p-1} - 1)
\eenn
for all $\pi < x < 2\pi$, hence $F'_p(x) < 0$.

\end{itemize}

\noindent
This proves that $0 < F_p(x) < 1$ for all $x \in (0,\pi) \cup (\pi,2\pi)$, from where the main claim immediately follows. \qed

\bigskip

\section{Measures, zeros and eigenvectors} 

Identification of the saturating states rests upon several technical results, which we gather in this appendix.

\blm\label{lm:intXf}
Let $X$ be topological space, $\mu$ a Borel measure on $X$ and $f : X \to [0,\infty)$ a continuous nonnegative function, with a nonempty set of zeros $Z = f^{-1}(\{0\})$. If
\be
\int_X f \, \df\mu = 0
\ee
then $\supp \mu \subseteq Z$. In particular, if $\mu(X) = 1$ and the set of zeros $Z$ is finite, then the measure has a form $\mu = \sum_{z \in Z} a_z \delta_z$ with some nonnegative constants $a_z$, such that $\sum_{z \in Z} a_z = 1$.
\elm

\noindent
\emph{Proof}. Let $B = \{ x \in X : \, f(x) > 0 \}$ and $A_n = \{ x \in X : \, f(x) \ge 1/n \}$ for all $n \in \nn$. By assumption, $f$ is a continuous function, hence $B$ is an open set and all $A_n$ are closed sets, which implies that all these sets are Borel. As 
\benn
f(x) 1_{A_n}(x) \ge \frac{1}{n} \, 1_{A_n}(x)
\eenn
holds for all $x \in X$, integration gives
\benn
0 = \int_X f \, \df\mu \ge \int_{A_n} f \, \df\mu \ge \frac{1}{n} \, \mu(A_n) ,
\eenn
hence $\mu(A_n) = 0$ for all $n$. Furthermore,
\benn
B = \bigcup_{n=1}^\infty A_n ,
\eenn
so by the countable subadditivity of the measure,
\benn
\mu(B) = \mu\left( \bigcup_{n=1}^\infty A_n \right) \le \sum_{n=1}^\infty \mu(A_n) = 0 .
\eenn
Therefore, $B$ is an open set of measure zero, which implies $B \cap \supp\mu = \0$ and $\supp\mu \subseteq Z$. Finite part of the claim follows immediately. \qed

\smallskip

The following lemma reconstructs form of a vector with the associated spectral measure which is a sum of Dirac measures.

\blm\label{lm:umudelta}
Let $\HH$ be a complex separable Hilbert space and let $A : D(A) \to \HH$ be a densely defined self-adjoint operator. Furthermore, let $\{\lambda_1,\dots,\lambda_n\}$ be a set of distinct real numbers and let $\{c_1,\dots,c_n\}$ be a set of nonzero complex numbers, such that $\sum_{i=1}^n |c_i|^2 = 1$. Then for each unit vector $u \in D(A)$ the associated spectral measure has a form $\mu_u = \sum_{i=1}^n |c_i|^2 \, \delta_{\lambda_i}$ if and only if there are orthonormal eigenvectors $e_i \in \Ker(A - \lambda_i I)$ such that $u = \sum_{i=1}^n c_i e_i$.
\elm

\noindent
\emph{Proof}.

\noindent
[$\Rightarrow$] \ Suppose that $\mu_u = \sum_{i=1}^n |c_i|^2 \, \delta_{\lambda_i}$. Let $e_i \defeq P_A(\{\lambda_i\}) u/c_i$ for each $i \in \{1,\dots,n\}$. Identity $\Ran\,P_A(\{\lambda_i\}) = \Ker(A - \lambda_i I)$ (see Problem 3.10 in \cite{Teschl}) implies that $e_i \in \Ker(A - \lambda_i I)$. As all these eigenvectors have pairwise different eigenvalues, they are orthogonal. Furthermore, each $e_i$ is a unit vector as
\benn
\norm{e_i}^2 = \norm{P_A(\{\lambda_i\}) u}^2/|c_i|^2 = \mu_u(\{\lambda_i\})/|c_i|^2 = 1 .
\eenn
For the Borel set $B = \rr - \{\lambda_1,\dots,\lambda_n\}$ we have $\norm{P_A(B) u}^2 = \mu_u(B) = 0$, hence $P_A(B)u = 0$. This gives us
\benn
u = Iu = P_A(\rr)u = \sum_{i=1}^n P_A(\{\lambda_i\})u = \sum_{i=1}^n c_i e_i .
\eenn

\medskip

\noindent
[$\Leftarrow$] \ Suppose that $u = \sum_{i=1}^n c_i e_i$. First, for each $e_i$ we have
\benn
0 = \norm{(A-\lambda_i I)e_i}^2 = \int_{\sigma(A)} (\lambda - \lambda_i)^2 \, \df\mu_{e_i}(\lambda) .
\eenn
As $f(\lambda) = (\lambda - \lambda_i)^2$ is a nonegative function with a single zero $\lambda_0 = \lambda_i$, Lemma \ref{lm:intXf} implies that $\mu_{e_i} = \delta_{\lambda_i}$. Let $P_A$ be a projection-valued spectral measure associated with the operator $A$. If $B \subseteq \rr$ is a Borel set such that $\lambda_i \notin B$, then
\benn
\norm{P_A(B) e_i}^2 = \mu_{e_i}(B) = \delta_{\lambda_i}(B) = 0 ,
\eenn
that is $P_A(B) e_i = 0$. On the other hand, if $\lambda_i \in B$, then
\benn
\norm{(I - P_A(B)) e_i}^2 = \norm{P_A(\rr - B) e_i}^2 = \mu_{e_i}(\rr - B) = \delta_{\lambda_i}(\rr - B) = 0 ,
\eenn
that is $P_A(B) e_i = e_i$. For any $v \in D(A)$, any $c \in \cc^\times$ and any Borel set $B \subseteq \rr$ we have
\benn
\mu_{c v}(B) = \lb cv, P_A(B) cv \rb = |c|^2 \mu_v(B) ,
\eenn
while for any pair of orthogonal eigenvectors $e_i$ and $e_j$
\begin{align*}
\mu_{e_i + e_j}(B) & = \lb e_i, P_A(B) e_i \rb + \lb e_j, P_A(B) e_j \rb + \lb e_i, P_A(B) e_j \rb + \lb e_j, P_A(B) e_i \rb \\
 & = \mu_{e_i}(B) + \mu_{e_j}(B) .
\end{align*}
The claim immediately follows from here. \qed

\medskip

Result of the theorem above has to be taken with the grain of salt if any of the eigenspaces has $\dim\Ker(A - \lambda_i I) \ge 2$. Namely, with the given vector $u$ and the associated spectral measure $\mu_u = \sum_{i=1}^n |c_i|^2 \, \delta_{\lambda_i}$ we have a canonical choice of eigenvectors, $e_i \defeq P_A(\{\lambda_i\}) u/c_i$. However, if we are only given a spectral measure, we encounter an ambiguity. For example, if $\dim\Ker(A - \lambda I) \ge m$ and $\{e_1,\dots,e_m\}$ are orthonormal vectors from $\Ker(H - \lambda I)$, then all vectors $u = \sum_{i=1}^m c_i e_i$ with $\sum_{i=1}^m |c_i|^2 = 1$, have the same associated spectral measure, $\mu_u = \delta_\lambda$.

\smallskip 

Another lemma, essential for the saturation of the MT bound, is bit more subtle.

\blm\label{lm:mumu}
Let $f : \rr \to [0,\infty)$ be a continuous nonnegative even function with a set of zeros $Z = \{-z,0,z\}$, where $z > 0$. Furthermore, let $\mu$ be a Borel measure on $\rr$, such that $\mu(\rr) = 1$ and
\be
\int_{\rr^2} f(x-y) \, \df(\mu \otimes \mu)(x,y) = 0 .
\ee
Then the measure has form $\mu = c \delta_a + (1-c) \delta_{a+z}$, with $0 \le c \le 1$ and $a \in \rr$.
\elm

\noindent
\emph{Proof}. Let $g : \rr^2 \to \rr$, $g(x,y) = x-y$, and $\nu = g_*(\mu \otimes \mu)$, that is $\nu(B) = (\mu \otimes \mu)(g^{-1}(B))$ for a Borel set $B \subseteq \rr$. Then the integral condition becomes
\benn
0 = \int_{\rr^2} (f \circ g)(x,y) \, \df(\mu \otimes \mu)(x,y) = \int_\rr f(s) \, \df\nu(s) = 0
\eenn
By Lemma \ref{lm:intXf}, we know that $\supp\nu \subseteq Z$. If $\supp\mu$ is a singleton, the main claim immediately follows (with $c=1$), so let us suppose that it contains at least two elements. We claim that $|a-b| \in Z$ for any two distinct $a,b \in \supp\mu$. Suppose, on the contrary, that $|a-b| \ne z$. Let $\delta = \min\{|a-b+z|,|a-b|,|a-b-z|\}/4$, $U = (a-\delta,a+\delta)$ and $V = (b-\delta,b+\delta)$. Then for any $x \in U$ and $y \in V$
\benn
|(x-y) - (a-b)| \le |x-a| + |y-b| < 2\delta ,
\eenn
from where it follows that $|x-y| \notin Z$. Therefore, $U \times V \subseteq g^{-1}(\rr - Z)$ and
\benn
0 = \nu(\rr - Z) = (\mu \otimes \mu)(g^{-1}(\rr - Z)) \ge (\mu \otimes \mu)(U \times V) = \mu(U)\mu(V) > 0,
\eenn
which is a contradiction. From the condition $|a-b| \in Z$ it is not difficult to see that $\supp\mu$ cannot contain more than two elements, which proves the main claim. \qed

\bigskip

\section{Fundamentals of Cantor measure} 

A family of (generalized) Cantor sets $\Can_\theta$ with the parameter $0 < \theta < 1/2$ is constructed by an iterated procedure. We begin with a partition of the set $C_0 \defeq [0,1]$ into $[0,\theta]$, $(\theta,1-\theta)$ and $[1-\theta,1]$, then we remove the middle interval, define set $C_1 \defeq [0,\theta] \cup [1-\theta,1]$, and repeat the procedure for each of the two components of $C_1$ with the same ratio of partitions. Once we have the sequence $(C_n)$ of sets, we define $\Can_\theta \defeq \bigcap_{n=0}^\infty C_n$, a compact, zero Lebesgue measure subset of $[0,1]$. Classical (ternary) Cantor set is $\Can_{1/3}$. Next we define Cantor function $F_\theta : [0,1] \to [0,1]$ as a pointwise limit of continuous nondecreasing functions $f_n : [0,1] \to [0,1]$, where $f_0(x) = x$ and
\be
f_{n+1}(x) = \left\{ \begin{array}{rl} f_n(x/\theta)/2 , & 0 \le x \le \theta \\[0.5em] 1/2 , & \theta \le x \le 1-\theta \\[0.5em] \big( 1 + f_n((x-1+\theta)/\theta) \big)/2 , & 1-\theta \le x \le 1 \end{array} \right.
\ee
for all $n \in \nn$. Bound
\be
\max_{x \in [0,1]} |f_{n+1}(x) - f_n(x)| \le 2^{-n} \max_{x \in [0,1]} |f_1(x) - f_0(x)| ,
\ee
assures that the sequence $(f_n)$ converges uniformly on $[0,1]$, so that $F_\theta$ is a nondecreasing continuous function as well. Moreover, $F'_\theta = 0$ a.e.~in $[0,1]$. Cantor function $F_\theta$ is a member of a wider family of functions sometimes referred to as ``Devil's staircase'' (for more details see \cite{Axler,SK}).

\smallskip

Finally, we define Cantor measure on $[0,1]$, starting from $\kappa_\theta((0,x]) = F_\theta(x)$ for all $x \in [0,1]$. Uniqueness of a Borel measure, based on nondecreasing continuous functions, is a standard result, see e.g.~Theorem A.3.~in \cite{Teschl}. Cantor measure $\kappa_\theta$ does not have any atoms, but it is supported on the zero Lebesgue measure Cantor set $\Can_\theta$, hence it is purely singular with respect to Lebesgue measure. Integration with respect to $\kappa_\theta$ may be streamlined by exploiting self-similarity of the measure,
\be
\kappa_\theta = \frac{1}{2} \left( \kappa_\theta \circ S^{-1} + \kappa_\theta \circ T^{-1} \right) ,
\ee
where $S(x) = \theta x$ and $T(x) = 1 - \theta + \theta x$. Then for every measureable function $f : [0,1] \to \rr$ we have (see e.g.~Theorem A.31 in \cite{Teschl})
\begin{align}
\int_{[0,1]} f \, \df\kappa_\theta & = \frac{1}{2} \left( \int_{[0,\theta]} f \, \df(S_* \kappa_\theta) + \int_{[1-\theta,1]} f \, \df(T_* \kappa_\theta) \right) \nonumber\\
 & = \frac{1}{2} \left( \int_{[0,1]} f(S(x)) \, \df\kappa_\theta(x) + \int_{[0,1]} f(T(x)) \, \df\kappa_\theta(x) \right) .
\end{align}
For example, if we introduce moments of the measure $\kappa_\theta$
\be\label{eq:momkappa}
M_n(\kappa_\theta) \defeq \int_{[0,1]} x^n \, \df\kappa_\theta(x) ,
\ee
we have $M_0(\kappa_\theta) = \kappa_\theta([0,1]) = 1$ and a recursive relation
\begin{align}
M_n(\kappa_\theta) & = \frac{1}{2} \int_{[0,1]} (\theta x)^n \, \df\kappa_\theta(x) + \frac{1}{2} \int_{[0,1]} (1 - \theta + \theta x)^n \, \df\kappa_\theta(x) \nonumber\\
 & = \frac{1}{2} \, \theta^n M_n(\kappa_\theta) + \frac{1}{2} \left( \theta^n M_n(\kappa_\theta) + \sum_{k=0}^{n-1} \binom{n}{k} (1-\theta)^{n-k} \theta^k M_k(\kappa_\theta) \right)
\end{align}
for all $n \in \nn$, so that
\be
M_n(\kappa_\theta) = \frac{1}{2(1-\theta^n)} \sum_{k=0}^{n-1} \binom{n}{k} (1-\theta)^{n-k} \theta^k M_k(\kappa_\theta) .
\ee
Similarly, assuming that we extend trivially $\kappa_\theta$ to the whole $\rr$, via $\kappa_\theta(\rr-\Can_\theta) = 0$, we can evaluate its Fourier transform, 
\begin{align}
\hat{\kappa}_\theta(k) & = \frac{1}{\sqrt{2\pi}} \int_{-\infty}^\infty e^{-ikx} \, \df\kappa_\theta(x) = \frac{1}{2} \, \big( 1 + e^{-i(1-\theta)k} \big) \hat{\kappa}_\theta(\theta k) \nonumber\\
 & = e^{-i(1-\theta)k/2} \cos((1-\theta)k/2) \hat{\kappa}_\theta(\theta k).
\end{align}
By iterating this relation we obtain
\be
\hat{\kappa}_\theta(k) = \left( \prod_{m=0}^{n-1} e^{-i(1-\theta)\theta^m k/2} \cos((1-\theta)\theta^m k/2) \right) \hat{\kappa}_\theta(\theta^n k) .
\ee
We have $\lim_{n\to\infty} \theta^n k = 0$, so using $\hat{\kappa}_\theta(0) = 1/\sqrt{2\pi}$ and the elementary sum $\sum_{m=0}^\infty (1-\theta) \theta^m = 1$, it follows that 
\be
\hat{\kappa}_\theta(k) = \frac{e^{-ik/2} }{\sqrt{2\pi}} \prod_{m=0}^\infty \cos((1-\theta)\theta^m k/2) .
\ee

\bigskip

\bibliographystyle{unsrt}
\bibliography{smqsl}

\end{document}